\documentclass[11pt]{article}
\usepackage{graphicx}
\usepackage{amssymb}
\usepackage{graphics}
\newcommand{\text}{\mbox}
\newcommand{\func}{\mbox}
\begin{document}
\parindent 0mm 
\setlength{\parskip}{\baselineskip} 
\thispagestyle{empty}
\pagenumbering{arabic} 
\setcounter{page}{0}
\mbox{ }
\rightline{UCT-TP-285/11}
\newline
\rightline{April 2011}
\newline
\vspace{0.2cm}
\begin{center}

{\Large {\bf Observability of an induced electric dipole moment of the
neutron from nonlinear QED}}

\end{center}

\vspace{.1cm}
\begin{center}
{\bf O. Zimmer}$^{(a)}$,
{\bf C. A. Dominguez}$^{(b),(c), (d)}$,  {\bf H. Falomir}$^{(e)}$,
{\bf M. Loewe}$^{(f)}$
\end{center}
\begin{center}
$^{(a)}$ Institut Laue Langevin, 38042 Grenoble, France

$^{(b)}$Centre for Theoretical \& Mathematical Physics,University of
Cape Town, Rondebosch 7700, South Africa

$^{(c)}$ Department of Physics, Stellenbosch University, Matieland 7600, South Africa

${(d)}$ National Institute of Theoretical Physics, Private Bag X1, Matieland 7602, South Africa

${(e)}$ Instituto de Fisica La Plata, Consejo Nacional de Ciencia y Tecnica, Facultad de Ciencias Exactas, Universidad Nacional de La Plata, Argentina

$^{(f)}$Facultad de F\'{i}sica, Pontificia Universidad Cat\'{o}lica de Chile, Casilla 306, Santiago 22, Chile

\end{center}
\vspace{0.3cm}
\begin{abstract}
\noindent
It has been shown recently that a neutron placed in an external quasistatic
electric field develops an induced electric dipole moment $\mathbf{p}_{%
\mathrm{IND}}$ due to quantum fluctuations in the QED vacuum. A feasible
experiment which could detect such an effect is proposed and described here.
It is shown that the peculiar angular dependence of $\mathbf{p}_{\mathrm{IND}%
}$ on the orientation of the neutron spin leads to a characteristic
asymmetry in polarized neutron scattering by heavy nuclei. This asymmetry can be 
of the order of $10^{-3}$ for neutrons with epithermal energies. For
thermalized neutrons from a hot moderator one still expects experimentally 
accessible values of the order of $10^{-4}$. The contribution of the induced effect to the neutron
scattering length is expected to be only one order of magnitude smaller than
that due to the neutron polarizability from its quark substructure. The
experimental observation of this scattering asymmetry would be the first ever
signal of nonlinearity in electrodynamics due to quantum fluctuations in
the QED vacuum.\\
\vspace{2cm}

\end{abstract}
\newpage
\section{Introduction}

Classical electrodynamics is well known to be a linear theory leading to
the superposition principle. At the quantum level the basic QED Lagrangian
remains quadratic in the electromagnetic fields, so that the theory still
appears to be linear. However, quantum fluctuations in the QED vacuum induce
nonlinear effects that lead to a breakdown of the superposition principle 
\cite{Heisenberg/1936}. In particular, these QED fluctuations make the
vacuum appear as an electrically and magnetically polarizable medium. The
size of these corrections in nonlinear QED (NLQED) is very tiny, so that
experiments with ultra-high intensity lasers have been proposed to search
for these effects, e.g. $\mathrm{e}^{+}\mathrm{e}^{-}$ pair production from
the vacuum \cite{Schwinger/1951}-\cite{Di Piazza/2009}, vacuum birefringence \cite
{Heinzl/2006}-\cite{Adler/2007}, light diffraction by a strong standing
electromagnetic wave \cite{Di Piazza/2008}, and nonlinear Compton
scattering \cite{Fried/1966}. A different proposal, involving quasistatic
external electromagnetic fields interacting with given electric or magnetic
sources, has been made recently \cite{Dominguez/2009/1}-\cite{Dominguez/2009}. In 
\cite{Dominguez/2009/1} general expressions were obtained for the induced
electric and magnetic fields in such circumstances, and applied to the case
of an electrically charged sphere in the presence of an external,
quasistatic magnetic field. As a result of QED nonlinearity there appears
an induced magnetic dipole moment, as well as corrections to the Coulomb
field of the sphere. In spite of this being a dramatic effect, experimental
detection appears very challenging. The complementary case of a purely
magnetic dipole moment placed in an external, quasistatic electric field ${%
\mathbf{E}_{0}}$ was considered in \cite{Dominguez/2009}. The result is an
induced electric dipole moment ${\mathbf{p}}_{\mathrm{IND}}$, plus
corrections to the magnetic field produced by the magnetic dipole. It was
then suggested that the neutron could be used as a probe in the presence of
large electric fields of order $|{\mathbf{E}_{0}}|\simeq 10^{10}$ V/m, such
as present in certain crystals. A distinctive feature of this induced
electric dipole moment, which should help in its detection, is its peculiar
dependence on the angle between ${\mathbf{p}}_{\mathrm{IND}}$ and ${\mathbf{E%
}_{0}}$, or equivalently the angle between ${\mathbf{p}}_{\mathrm{IND}}$ and
the neutron spin.\\
In this paper we follow up on the experimental observability of such an
induced electric dipole moment of the neutron. On the theoretical side we
complete the analysis of \cite{Dominguez/2009} by computing the interaction
Hamiltonian of the neutron immersed in a large external quasistatic electric
field ${\mathbf{E}_{0}}$, and an external, quasistatic, magnetic field ${%
\mathbf{B}_{0}}$ of ordinary strength. Given the nonlinearity of the
problem one needs to check that (a) the magnetic interaction energy is of
the usual form, (nonlinear magnetic corrections due to ${\mathbf{B}_{0}}$
are negligible), and (b) that the induced electric dipole does interact with
the electric field ${\mathbf{E}_{0}}$ that generates it. The latter
interaction energy is expected to have the standard functional form $H_{%
\mathrm{int}}\propto {\mathbf{p}}_{\mathrm{IND}}\cdot {\mathbf{E}_{0}}$,
albeit with an a-priori unknown coefficient which we determine. Next, we
study the quantum behaviour of ${\mathbf{p}}_{\mathrm{IND}}$ by means of the
Heisenberg equation of motion. This is important for experiments based on
potential changes in the Larmor frequency of the neutron spin around an
external magnetic field due to the presence of ${\mathbf{p}}_{{\mathrm{IND}}}
$. We find no effect here, thus ruling out experiments of this type to
detect an induced electric dipole moment of the neutron. Finally, we discuss
in some detail a different approach based on neutron-nucleus scattering and
conclude that this experiment offers an excellent opportunity to observe
such an effect. This is due to the peculiar angular dependence of ${\mathbf{p%
}}_{\mathrm{IND}}$. We find that for sufficiently large momentum transfers,
a scattering asymmetry is induced with such particular characteristics that
it would be easy to distinguish from other standard effects. The
experimental discovery of such an asymmetry would be the first ever signal
of a nonlinear effect in electrodynamics due to quantum fluctuations in
the QED vacuum.
\section{Induced electric dipole moment of the neutron}
An appropriate framework to discuss nonlinear effects induced by quantum
fluctuations in the QED vacuum is that of the Euler-Heisenberg Lagrangian 
\cite{Heisenberg/1936}. This is obtained from the weak field asymptotic
expansion of the QED effective action at one loop order leading to%
\begin{equation}
\mathcal{L}_{\mathrm{EH}}^{\left( 1\right) }=\zeta \left( 4\mathcal{F}^{2}+7%
\mathcal{G}^{2}\right) +...,  \label{L-EH}
\end{equation}%
where the omitted terms are of higher order in the expansion parameter $%
\zeta $. In  SI units
\begin{equation}
\zeta =\frac{2\alpha _{\mathrm{EM}}^{2}\epsilon _{0}^{2}\hbar ^{3}}{%
45m_{e}^{4}c^{5}}\simeq 1.3\times 10^{-52}\,\frac{\mathrm{Jm}}{\mathrm{V}^{4}%
},
\end{equation}%
with $\alpha _{\mathrm{EM}}=e^{2}/(4\pi \epsilon _{0}\hbar c)$ the
electromagnetic fine structure constant, $m_{\mathrm{e}}$ and $e$ are the
mass and charge of the electron, respectively, and $c$ the speed of light.
The invariants $\mathcal{F}$ and $\mathcal{G}$ are defined as
\begin{equation}
\mathcal{F}=\frac{1}{2}\left( \mathbf{E}^{2}-c^{2}\mathbf{B}^{2}\right) =-%
\frac{1}{4}F_{\mu \nu }F^{\mu \nu },
\end{equation}
\begin{equation}
\mathcal{G}=c\text{\/}\mathbf{E}\cdot \mathbf{B}=-\frac{1}{4}F_{\mu \nu }%
\widetilde{F}^{\mu \nu },
\end{equation}%
with $F_{\mu \nu }=\partial _{\mu }A_{\nu }-\partial _{\nu }A_{\mu }$ and $%
\widetilde{F}^{\mu \nu }=\frac{1}{2}\epsilon ^{\mu \nu \rho \sigma }F_{\rho
\sigma }$. The so-called critical field $E_{\mathrm{c}}$, which plays the
role of a reference field strength for the onset of nonlinearity, is given
by
\begin{equation}
E_{\mathrm{c}}=\frac{m_{\mathrm{e}}^{2}c^{3}}{\hbar e}\simeq 1.3\times
10^{18}\,\frac{\mathrm{V}}{\mathrm{m}}.  \label{E-crit}
\end{equation}%
This estimate is obtained by computing the electric field needed to produce
an electron-positron pair in a spatial length of one Compton wavelength. For
fields stronger than $E_{\mathrm{c}}$ the weak-field asymptotic expansion
leading to Eq. (\ref{L-EH}) breaks down. In \cite{Dominguez/2009/1} general
expressions were obtained for electric and magnetic fields induced by
nonlinearity, to leading order in $\zeta $, in the presence of external
quasistatic weak fields (smaller than $E_{\mathrm{c}}$) and arbitrary
sources. These induced fields are
\begin{equation}
\mathcal{E}\left( \mathbf{x}\right) =\frac{\zeta }{2\pi \epsilon _{0}^{2}}%
\nabla _{\mathbf{x}}\int \frac{\mathrm{d}^{3}y}{\left\vert \mathbf{x}-%
\mathbf{y}\right\vert }\nabla _{\mathbf{y}}\cdot \left( 4\mathcal{F}_{%
\mathrm{M}}\mathbf{D}_{\mathrm{M}}+\frac{7}{c}\mathcal{G}_{\mathrm{M}}%
\mathbf{H}_{\mathrm{M}}\right) ,  \label{E-ind}
\end{equation}
\begin{equation}
\mathcal{B}\left( \mathbf{x}\right) =\frac{\zeta }{2\pi \epsilon
_{0}^{2}c^{2}}\nabla _{\mathbf{x}}\times \int \frac{\mathrm{d}^{3}y}{%
\left\vert \mathbf{x}-\mathbf{y}\right\vert }\nabla _{\mathbf{y}}\times
\left(\frac{}{} -4\mathcal{F}_{\mathrm{M}}\mathbf{H}_{\mathrm{M}}+7c\mathcal{G}_{%
\mathrm{M}}\mathbf{D}_{\mathrm{M}}\frac{}{}\right) ,  \label{B-ind}
\end{equation}%
where $\mathbf{D}_{\mathrm{M}}$ and $\mathbf{H}_{\mathrm{M}}$ are the
Maxwell (classical) fields produced by the arbitrary sources. Notice that
these fields vanish as $\hbar \rightarrow 0$ ($\zeta \rightarrow 0$). In the
case of a current density uniformly distributed on the surface of a sphere
of radius $a$, or equivalently, for a uniformly magnetized sphere of the
same radius, the Maxwell, magnetic dipole type field, is given by
\begin{equation}
\mathbf{B}_{\mathrm{d}}=\frac{\mu _{0}}{4\pi }\left\{ \frac{3\left( \mathbf{m%
}\cdot \mathbf{e}_{r}\right) \mathbf{e}_{r}-\mathbf{m}}{r^{3}}\;\Theta \left(
r-a\right) +\frac{2\mathbf{m}}{a^{3}}\;\Theta \left( a-r\right) \right\} ,
\label{B-d}
\end{equation}
where $\mathbf{m}$ is identified with the magnetic dipole moment of the
source, and $\mathbf{e}_{r}$ is a unit vector in the radial direction. Since
the central expressions, Eqs.(\ref{E-ind}) and (\ref{B-ind}), were derived
assuming $E\equiv cB<E_{\mathrm{c}}$ the following constraint follows
\begin{equation}
\frac{\left\vert \mathbf{m}\right\vert }{a^{3}}<\frac{2\pi m_{\mathrm{e}%
}^{2}c^{2}}{\hbar e\mu _{0}}.  \label{a-constraint}
\end{equation}
For instance, if $|\mathbf{m}|=0.96\;\times \,10^{-26}\,\mathrm{A \,m}^{2}$, as
for the neutron, then it follows that $a\gtrsim 10\,\mathrm{fm}$. If this
magnetic source is placed in an external, constant electric field $\mathbf{E}%
_{0}$ it has been shown \cite{Dominguez/2009} that there is an induced
electric field of the dipole type
\begin{equation}
\mathcal{E}\left( \mathbf{x}\right) =-\nabla _{\mathbf{x}}\left[ \frac{1}{%
4\pi \epsilon _{0}}\frac{\mathbf{p}\left( \psi \right)_{\mathrm{IND}} \cdot 
\mathbf{e}_{r}}{\left\vert \mathbf{x}\right\vert ^{2}}\right] +\mathcal{O}%
\left( \left\vert \mathbf{x}\right\vert ^{-6}\right) .  \label{E-induced}
\end{equation}
where $\psi $ is the angle between the external electric field, lying in the 
$x$-$z$ plane, and the magnetic dipole moment pointing along the $z$ axis,
i.e. $\mathbf{E}_{0}=\left\vert \mathbf{E}_{0}\right\vert \left( \sin \psi \,%
{\mathbf{e}}_{x}+\cos \psi \,{\mathbf{e}}_{z}\right) $, and $\mathbf{m}=|%
\mathbf{m}|\mathbf{e}_{z}$. The induced electric dipole moment $\mathbf{p}(\psi )_{
\mathrm{IND}}$ is given by 
\begin{equation}
\mathbf{p}\left( \psi \right)_{\mathrm{IND}} =\frac{\zeta \mu _{0}\left\vert 
\mathbf{m}\right\vert ^{2}\left\vert \mathbf{E}_{0}\right\vert }{10\pi
\epsilon _{0}a^{3}}\left[ 36\frac{\mathbf{E}_{0}}{|\mathbf{E}_{0}|}-49\left( 
\frac{\mathbf{E}_{0}}{|\mathbf{E}_{0}|}\cdot {\mathbf{e}}_{x}\right) {%
\mathbf{e}}_{x}\right] \,.  \label{p-ind}
\end{equation}%
This induced electric field is of the electric dipole type in its radial $1/|%
\mathbf{x}|^{3}$ dependence, but it has a manifestly peculiar angular
dependence. For instance, along the $z$ axis, and unlike a standard electric
dipole field, it has a non-zero component along $\mathbf{e}_{\theta }$ that
depends on the azimuthal angle $\phi $. It also has a non-zero component
along the direction of $\mathbf{e}_{\phi }$, as may be appreciated by
writing the induced electric field in spherical coordinates ($r$, $\theta $, 
$\phi $), i.e.
\begin{eqnarray}
\mathcal{E}(\mathbf{x}) &=&\;\frac{\zeta \mu _{0}|\mathbf{m}|^{2}|\mathbf{E}%
_{0}|}{40\pi ^{2}\epsilon _{0}^{2}a^{3}\mathbf{|x|}^{3}} \left\{\frac{}{} 2\,\left[
36\cos \theta \cos \psi -13\sin \theta \cos \phi \sin \psi \right] \,{%
\mathbf{e}_{r}}\right.   \nonumber \\
&&+\left. \left[ 13\cos \theta \cos \phi \sin \psi +36\sin \theta \cos \psi %
\right] {\mathbf{e}_{\theta }}-13\,\sin \phi \sin \psi \;{\mathbf{e}_{\phi }}%
 \frac{}{} \right\} \;.
\end{eqnarray}%
In addition to the induced electric field Eq.(\ref{E-induced}), there is an
induced magnetic field (a correction to the field produced by the magnetic
dipole source), which can be derived from a vector potential, i.e. $\mathcal{%
B}(\mathbf{x})=\nabla \times \mathcal{A}(\mathbf{x})$, where after a lengthy
calculation one finds
\begin{eqnarray}
\mathcal{A}\left( \mathbf{x}\right)  &=&\frac{\zeta \mu _{0}}{4\pi \epsilon
_{0}\left\vert \mathbf{x}\right\vert ^{2}}\left\{\frac{}{} 4\left\vert \mathbf{E}%
_{0}\right\vert ^{2}\left( \mathbf{e}_{r}\times \mathbf{m}\right) -7\left[ 
\mathbf{m}\cdot \mathbf{E}_{0}+3\left( \mathbf{E}_{0}\cdot \mathbf{e}%
_{r}\right) \left( \mathbf{m}\cdot \mathbf{e}_{r}\right) \right] \left( 
\mathbf{e}_{r}\times \mathbf{E}_{0}\right) \right.   \label{A-calli} \\
&&\left. +7\left( \mathbf{E}_{0}\cdot \mathbf{e}_{r}\right) \left( \mathbf{m}%
\times \mathbf{E}_{0}\right) \frac{}{} \right\} \left[ 1+\mathcal{O}\left( \frac{\mu
_{0}\left\vert \mathbf{m}\right\vert ^{2}}{a^{6}\epsilon _{0}\left\vert 
\mathbf{E}_{0}\right\vert ^{2}}\right) \right] +\mathcal{O}\left( \left\vert 
\mathbf{x}\right\vert ^{-4}\right) .  \nonumber
\end{eqnarray}%
Notice that while $\mathcal{E}$ grows linearly with $|\bf{E}_{0}|$, $\mathcal{B}$
depends quadratically on $|\bf{E}_{0}|$.\\
We proceed to discuss the interaction energy of the magnetic dipole source
and its induced electric dipole with the external constant field $\mathbf{E}%
_{0}$, and with an external uniform magnetic field $\mathbf{B}_{0}$ weak
enough not to induce nonlinear effects, i.e. $c|\mathbf{B}_{0}|\ll E_{%
\mathrm{c}}$. Given the nonlinearity of the problem it is important to
verify that the magnetic interaction Hamiltonian has the expected form $-%
\mathbf{m}\cdot \mathbf{B}_{0}$, given the strength of $\mathbf{B}_{0}$. In
addition, the electric interaction energy of the induced electric dipole and
the external field $\mathbf{E}_{0}$ is a-priori unknown. This need not be
exactly of the form
\begin{equation}
H_{\mathrm{int}}=-\frac{1}{2}\mathbf{p}\cdot \mathbf{E}_{0}\mathbf{,}
\label{H-el-pol}
\end{equation}%
as one would obtain for a linearly polarizable particle immersed in an
external electric field, e.g. for a polarizable neutron on account of its
quark substructure. In fact, the electric interaction Hamiltonian due to
nonlinearity lacks the factor $1/2$ as shown next. The canonical
energy-momentum tensor is defined as
\begin{equation}
T_{\text{ }\nu }^{\mu }=\frac{\partial \mathcal{L}_{\mathrm{tot}}}{\partial
\left( \partial _{\mu }A_{\alpha }\right) }\left( \partial _{\nu }A_{\alpha
}\right) -\mathcal{L}_{\mathrm{tot}} \;\delta _{\text{ }\nu }^{\mu },
\end{equation}%
where the total Lagrangian density is $\mathcal{L}_{\mathrm{tot}}=\mathcal{L}%
-j_{\mu }A^{\mu }$,  with $\mathcal{L}=\epsilon _{0}\mathcal{F}+\mathcal{L}_{%
\mathrm{EH}}^{\left( 1\right) }$  and $\mathcal{L}_{\mathrm{EH}}^{\left(
1\right) }$ given in Eq.(\ref{L-EH}). This equation can be rewritten as
\begin{eqnarray}
T_{\text{ }\nu }^{\mu } &=&\left( \frac{\partial \mathcal{L}}{\partial 
\mathcal{F}}F^{\mu \alpha }+\frac{\partial \mathcal{L}}{\partial \mathcal{G}}%
\widetilde{F}^{\mu \alpha }\right) F_{\alpha \nu }- \mathcal{L} \,\delta _{\text{
}\nu }^{\mu }+\left( j\cdot A\right) \, \delta _{\text{ }\nu }^{\mu }  \nonumber
\\
&&+A_{\nu }\partial _{\alpha }\left( \frac{\partial \mathcal{L}}{\partial 
\mathcal{F}}F^{\mu \alpha }+\frac{\partial \mathcal{L}}{\partial \mathcal{G}}%
\widetilde{F}^{\mu \alpha }\right) -\partial _{\alpha }\left[ \left( \frac{%
\partial \mathcal{L}}{\partial \mathcal{F}}F^{\mu \alpha }+\frac{\partial 
\mathcal{L}}{\partial \mathcal{G}}\widetilde{F}^{\mu \alpha }\right) A_{\nu }%
\right] .
\end{eqnarray}%
Since the last term on the right hand side above is the total divergence of
an anti-symmetric tensor, employing the equations of motion
\begin{equation}
\partial _{\beta }\left( \frac{\partial \mathcal{L}}{\partial \mathcal{F}}%
F^{\beta \alpha }+\frac{\partial \mathcal{L}}{\partial \mathcal{G}}%
\widetilde{F}^{\beta \alpha }\right) =j^{\alpha },
\end{equation}%
one can define another energy-momentum tensor as
\begin{eqnarray}
\theta _{\text{ }\nu }^{\mu } &=&T_{\text{ }\nu }^{\mu }+\partial _{\alpha }%
\left[ \left( \frac{\partial \mathcal{L}}{\partial \mathcal{F}}F^{\mu \alpha
}+\frac{\partial \mathcal{L}}{\partial \mathcal{G}}\widetilde{F}^{\mu \alpha
}\right) A_{\nu }\right]   \nonumber \\
&=&\left( \frac{\partial \mathcal{L}}{\partial \mathcal{F}}F^{\mu \alpha }+%
\frac{\partial \mathcal{L}}{\partial \mathcal{G}}\widetilde{F}^{\mu \alpha
}\right) F_{\alpha \nu }-\mathcal{L}\, \delta _{\text{ }\nu }^{\mu }+\left(
j_{\alpha }A^{\alpha }\right) \delta _{\text{ }\nu }^{\mu }-j^{\mu }A_{\nu }.
\end{eqnarray}%
Notice that this tensor is symmetric and gauge invariant except for the last
two terms. The total energy density of the system is defined as the
component $\theta _{\text{ }0}^{0}$, 
\begin{equation}
\mathcal{H}_{\mathrm{tot}}=\theta _{\text{ }0}^{0}=\frac{\partial \mathcal{L}%
}{\partial \mathbf{E}}\cdot \mathbf{E}-\mathcal{L}-\mathbf{j}\cdot \mathbf{A}%
=\mathbf{D\cdot E}-\mathcal{L}-\mathbf{j}\cdot \mathbf{A}.
\end{equation}%
In general, for a given configuration of the fields the interaction
Hamiltonian is defined as the difference of the total Hamiltonian with and
without the sources. In a quantum theory it is defined as the difference of
the total Hamiltonian evaluated at the fields in the interaction picture,
with and without the external sources. Then, the interaction Hamiltonian $H_{%
\mathrm{int}}$, i.e. the volume integral of the interaction Hamiltonian
density ${\mathcal{H}}_{\mathrm{int}}$ is
\begin{equation}
H_{\mathrm{int}}=\int \mathcal{H}_{\mathrm{int}} \, \mathrm{d}^{3}r=-\int 
\mathbf{j}\cdot \mathbf{A} \, \mathrm{d}^{3}r=-\int \mathbf{j}\cdot \left( 
\mathbf{A}_{0}+\mathcal{A}\right) \, \mathrm{d}^{3}r,  \label{H-int}
\end{equation}%
where $\mathbf{A}=\mathbf{A}_{0}+\mathcal{A}$, with $\mathcal{A}$ given in
Eq.(\ref{A-calli}), and $\mathbf{A}_{0}$ is the vector potential associated
with $\mathbf{B}_{0}$, i.e. $\mathbf{B}_{0}=\nabla \times \mathbf{A}_{0}$,
and $\mathbf{A}_{0}=\frac{1}{2}\,\mathbf{B}_{0}\times \mathbf{r}$. The
current $\mathbf{j}$ corresponding to the magnetized sphere producing the
field, Eq.(\ref{B-d}), is $\mathbf{j}=\frac{3}{4\pi a^{3}}\,\mathbf{m}\times 
\mathbf{e}_{r}\,\delta (r-a)$. In Eq.(\ref{H-int}) the self energy of the
magnetized sphere, independent of the external field, has been omitted.
After performing the integration in Eq.(\ref{H-int}) one finds
\begin{equation}
H_{\mathrm{int}}=\mathbf{-m\cdot B}_{0}\mathbf{-p}\left( \psi
\right)_{\mathrm{IND}} \mathbf{\cdot E}_{0},  \label{pot-energy}
\end{equation}%
which has the correct magnetic interaction term as in the linear theory. The
electric interaction energy is of the expected form, but it involves a
coefficient different from the case of linear QED as a result of
nonlinearity. In the absence of the external magnetic field $\mathbf{B}_{0}$%
, and using Eq.(\ref{p-ind}), the interaction Hamiltonian becomes
\begin{equation}
H_{\mathrm{int}}=-\frac{\zeta \mu _{0}\left\vert \mathbf{m}\right\vert
^{2}\left\vert \mathbf{E}_{0}\right\vert ^{2}}{10\pi \epsilon _{0}a^{3}}%
\left( 36-49\sin ^{2}\psi \right) =\frac{\zeta \mu _{0}\left\vert \mathbf{m}%
\right\vert ^{2}\left\vert \mathbf{E}_{0}\right\vert ^{2}}{10\pi \epsilon
_{0}a^{3}}\left( 13-49\cos ^{2}\psi \right) .  \label{Hamiltonian}
\end{equation}%
Notice the dependence of $H_{\mathrm{int}}$ on $a^{-3}$. It should be
pointed out that in an experimental situation one would typically be
interested in a point magnetic dipole.  This source would produce very strong fields
in its proximity so that the limit $a\rightarrow 0$  would obviously not be allowed. Instead,  we assume that even in such a case the
large distance solution for the fields is well described by the first order
approximation to the effective Lagrangian $\mathcal{L}_{\mathrm{EH}}^{\left(
1\right) }$ in Eq.(\ref{L-EH}). We also assume that this solution is robust
against short distance modifications of the source as long as its symmetry
is preserved. In this sense the parameter $a$ is to be considered as a
measure of our ignorance about the higher order corrections to this
effective Lagrangian, something necessary when dealing with strong fields. The specific value of $a$
will be discussed later in Section 4. The fact that $H_{\mathrm{int}}$ depends on the orientation of $\mathbf{m}$ with
respect to $\mathbf{E}_{0}$ through the angle $\psi$ can be used as
a distinctive feature in the design of an experimental asymmetry as described below in Section 5.\\
We consider next the quantum behaviour of ${\mathbf{p}}_{\mathrm{\mathrm{IND}}%
}$ using the Heisenberg equation of motion. To this end we consider a
particle with magnetic dipole moment ${\mathbf{m}}$  related to the spin
through the standard relation ${\mathbf{m}}= g \hbar {\mathbf{S}}$,
where $g$ is the gyromagnetic ratio. Assuming that the dynamics of
this particle is described by the Hamiltonian Eq.(\ref{pot-energy}), and
given that ${\mathbf{p}}_{\mathrm{IND}}\propto |{\mathbf{m}}|^{2}$, $H_{%
\mathrm{int}}$ to first order in $\zeta $ contains only quadratic terms in
the spin, whose components are the dynamical variables of the problem. The
effective Hamiltonian involving these dynamical variables must be
symmetrized in order to ensure Hermiticity. Hence, the quadratic terms in
the spin entering the Heisenberg equation of motion lead to the commutator
\begin{equation}
\left[ \left\{ S_{i},S_{j}\right\} ,S_{k}\right] =i\epsilon _{jkl}\left\{
S_{i},S_{l}\right\} +i\epsilon _{ikl}\left\{ S_{j},S_{l}\right\} .
\end{equation}%
For a spin $1/2$ particle, such as the neutron, we have $S_{i}=\frac{1}{2}%
\sigma _{i}$, and $\left\{ S_{i},S_{l}\right\} =\frac{1}{2}\delta _{il}$. In
this case,
\begin{equation}
\left[ \left\{ S_{i},S_{j}\right\} ,S_{k}\right] =i\epsilon _{jkl}\frac{1}{2}%
\delta _{il}+i\epsilon _{ikl}\frac{1}{2}\delta _{jl}=\frac{i}{2}\left(
\epsilon _{jki}+\epsilon _{ikj}\right) =0.
\end{equation}%
Therefore, $d\mathbf{S}/dt=0$ so that if one is interested in the time
evolution of a spin $1/2$ particle, and Eq.(\ref{pot-energy}) describes its
effective Hamiltonian, we find no contribution from this leading order
nonlinear correction. In other words, the precession of the spin is not
affected. This is not the case, though, for spin-one particles. This
unfortunate feature rules out experiments to detect the induced electric
dipole moment of the neutron based on Larmor frequency changes. A different
approach involving neutron scattering off nuclei is discussed next.
\section{Neutron-atom scattering amplitude and cross section}
Scattering of slow neutrons by a free atom can be described by a scattering
amplitude in the Born-approximation, which in the center of mass system is given
by (see e.g. \cite{Turchin/1965})
\begin{equation}
f\left( \mathbf{q},\mathbf{s}\right) =-\frac{M}{2\pi \hbar ^{2}}\int \exp
\left( i\mathbf{q\cdot r}\right) H_{\mathrm{int}}\left( \mathbf{q},\mathbf{s}%
\right) \mathrm{d}^{3}r.  \label{f}
\end{equation}%
where $M$ is the reduced mass
\begin{equation}
M=\frac{m_{\mathrm{n}}m_{\mathrm{A}}}{m_{\mathrm{n}}+m_{\mathrm{A}}},
\label{reduced mass}
\end{equation}%
with $m_{\mathrm{n}}$ the neutron mass, and $m_{\mathrm{A}}$ the mass of the
atom. The three-momentum transfer $\mathbf{q}$ is 
$\mathbf{q}=\mathbf{k-k}^{\prime}$, with  
  $\mathbf{k}$ and $\mathbf{k}^{\prime
}$  the neutron wave vectors before and after scattering, respectively, and $\mathbf{s}$
\textbf{\ }is the neutron spin in units of $\hbar $. The magnitude of $\bf{q}$ will be denoted as
$|\mathbf{q}| \equiv q$ in the sequel.
The total Hamiltonian $H_{\mathrm{int}}$ involves all known interactions
between the neutron and the atom, to which we add now the new interaction
due to NLQED given in Eq.(\ref{Hamiltonian}). Correspondingly, the total
scattering amplitude can be written as 
\begin{equation}
f\left( \mathbf{q},\mathbf{s}\right) =f_{\mathrm{N}}\left( \mathbf{q},%
\mathbf{s}\right) +f_{\mathrm{MAG}}\left( \mathbf{q},\mathbf{s}\right) +f_{%
\mathrm{e}}\left( \mathbf{q}\right) +f_{\mathrm{POL}}\left( \mathbf{q}%
\right) +f_{\mathrm{SO}}\left( \mathbf{q},\mathbf{s}\right) +f_{\mathrm{PV}%
}\left( \mathbf{q},\mathbf{s}\right) +f_{\mathrm{IND}}\left( \mathbf{q},%
\mathbf{s}\right) ,  \label{scattering amp}
\end{equation}%
where the various contributions are as follows. The term $f_{\mathrm{N}%
}\left( \mathbf{q},\mathbf{s}\right) $ is due to the hadronic interaction of
the neutron with the nucleus, and is usually the dominant term. The
amplitude $f_{\mathrm{MAG}}\left( \mathbf{q},\mathbf{s}\right) $ corresponds
to the interaction of the neutron magnetic moment with the atomic magnetic
field (for atoms with unpaired electrons). This term can be of a similar size
as $f_{\mathrm{N}}\left( \mathbf{q},\mathbf{s}\right) $. The next three
terms arise from various electromagnetic interactions, i.e. $f_{%
\mathrm{e}}\left( \mathbf{q}\right) $ is due to scattering of the neutron
charge radius by the electric charges in the atom, $f_{\mathrm{POL}}\left( 
\mathbf{q}\right) $ arises from the electric polarizability of the neutron
due to its quark substructure, and $f_{\mathrm{SO}}\left( \mathbf{q},\mathbf{%
s}\right) $ corresponds to the spin-orbit interaction of the neutron in the
electric field of the nucleus. The term $f_{\mathrm{PV}}\left( \mathbf{q},%
\mathbf{s}\right) $ is a weak interaction, parity-violating amplitude which
we list separately from $f_{\mathrm{N}}\left( \mathbf{q},\mathbf{s}\right) $
as it has a different dependence on neutron spin. Finally, $f_{\mathrm{IND}}$
is the new component due to the induced electric dipole moment of the
neutron, which we wish to isolate experimentally.\\
The scattering amplitude, Eq.(\ref{scattering amp}), enters the differential
cross section for elastic neutron scattering by a single atom in the ground
state,
\begin{equation}
\frac{\mathrm{d}\sigma }{\mathrm{d}\Omega }\left( \mathbf{q},\mathbf{P}%
\right) =\left\langle \left\vert f\left( \mathbf{q},\mathbf{s}\right)
\right\vert ^{2}\right\rangle ,  \label{sigma}
\end{equation}%
which includes an ensemble average over nuclear and electronic spin degrees
of freedom (if present), and the neutron spin. The incident neutrons are
characterized by a polarization defined as $\mathbf{P}=2\left\langle \mathbf{%
s}\right\rangle $. In the absence of nuclear and electronic polarization of the
atom, the case of interest here, the kinematic   scattering variables are $%
\mathbf{q}$ and $\mathbf{P}$. Experimentally, one determines neutron
scattering cross sections using a sample containing a macroscopic number of
atoms. Considering a single atomic species, the ensemble average in Eq.(\ref%
{sigma}) still has to account for the isotopic composition and the different
states of total spin of a neutron scattering off a nucleus with non-zero
spin. 
For slow neutrons with wavelengths much larger than the nuclear radius $R_{
\mathrm{N}}$ the hadronic amplitude $f_{\mathrm{N}}$ is practically independent of 
$\mathbf{q}$. This is in the absence of nuclear resonances for thermal
and epithermal neutrons, i.e. for the energy range of interest here.
Scattering thus proceeds in an s-wave and is isotropic in the
center of mass system. One defines a neutron scattering length operator as
\begin{equation}
a_{\mathrm{N}}\left( \mathbf{s}\right) =-\lim_{q\rightarrow 0}f_{\mathrm{N}%
}\left( \mathbf{q},\mathbf{s}\right) .  \label{scattering length}
\end{equation}%
For a nucleus with spin $\hbar \mathbf{I}$ one has
\begin{equation}
a_{\mathrm{N}}\left( \mathbf{s}\right) =\frac{\left( I+1\right) a_{+}+Ia_{-}%
}{2I+1}+\frac{2\left( a_{+}-a_{-}\right) }{2I+1}\mathbf{s}\cdot \mathbf{I},
\label{a-spin}
\end{equation}%
where $a_{+}$ and $a_{-}$ are the eigenvalues of $a_{\mathrm{N}}\left( 
\mathbf{s}\right) $ for the two states of total spin $I  \pm 1/2$ (see e.g. 
\cite{Turchin/1965}). For a sample with all nuclear species unpolarized,
scattering by the $i$th isotope enters with statistical weight factors $%
w_{i+}=\left( I_{i}+1\right) /\left( 2I_{i}+1\right) $ and $%
w_{i-}=I_{i}/\left( 2I_{i}+1\right) $. The leading term in the cross section
is then given by
\begin{equation}
\overline{\left\vert a_{\mathrm{N}}\right\vert ^{2}}=\sum_{i}c_{i}\left[
w_{i+}\left\vert a_{i+}\right\vert ^{2}+w_{i-}\left\vert a_{i-}\right\vert
^{2}\right] ,
\end{equation}%
where the bar indicates the averaging over isotopes and spin states, and $%
c_{i}$ stands for the relative abundance of the $i$th isotope. In
next-to-leading order the cross section contains interference terms between
small amplitudes like $f_{\mathrm{IND}}$ and a usually dominant coherent
nuclear scattering length $\overline{a_{\mathrm{N}}}$, which for unpolarized
nuclei is given by
\begin{equation}
\overline{a_{\mathrm{N}}}=\sum_{i}c_{i}\left[ w_{i+}a_{i+}+w_{i-}a_{i-}%
\right] .
\end{equation}%
Most scattering lengths $\overline{a_{\mathrm{N}}}$ are found to be positive
with typical values of a few $\mathrm{fm}$. Neutron optical measurements determine a coherent {\it bound scattering length} $\overline{b}$, related to the corresponding scattering length $\overline{a}$
of a free atom through 
\begin{equation}
\overline{b}=\overline{a}\left( 1+\frac{m_{\mathrm{n}}}{m_{\mathrm{A}}}%
\right) .
\end{equation}%
This relation includes the contributions $-\lim_{q\rightarrow 0}\overline{%
f_{i}}$ due to all amplitudes ($i=\mathrm{N},$ $\mathrm{POL}...$) appearing
in Eq.(\ref{scattering amp}). Lacking sufficiently accurate theoretical
predictions for the nuclear part, however, one cannot even extract from $%
\overline{b}$  the sum of all non-hadronic components,  which normally
contribute less than $1\%$. Instead, one needs to perform measurements for $%
q\neq 0$. Values for $\overline{b}$ and the total neutron scattering cross
section of an atom fixed in space, $\sigma _{\mathrm{s,b}}=4\pi \overline{
b}^{2}$, can be found  e.g. in \cite{Rauch/2002}. For later
use we quote the values for lead with natural isotopic abundances
\begin{eqnarray}
\overline{b} &=&\left( 9.401\pm 0.002\right) \,\mathrm{fm},
\label{values-Pb} \\
\sigma _{\mathrm{s,b}} &=&\left( 11.187\pm 0.007\right) \times 10^{-24}\,%
\mathrm{cm}^{2}.  \nonumber
\end{eqnarray}
For low neutron energies one also has to take into account interference
effects of the neutron waves scattered from different atoms, as e.g. in
condensed-matter studies. Classical examples are Bragg scattering by single
crystals and measurements of phonon dispersion relations. However, for
sufficiently large momentum transfer $ \mathbf{q}$ as considered
here, interatomic interferences and eventual nuclear spin correlations
between different atoms can be neglected. We thus consider the cross section
in the center of mass system as given by 
\begin{equation}
\frac{\mathrm{d}\sigma }{\mathrm{d}\Omega }\simeq \overline{\left\vert a_{%
\mathrm{N}}\right\vert ^{2}}+\left\langle \left\vert f_{\mathrm{SO}}\left( 
\mathbf{q},\mathbf{s}\right) \right\vert ^{2}\right\rangle
+...-\sum_{j}\left\langle 2\func{Re}\left[ \overline{a_{\mathrm{N}}}%
f_{j}\left( \mathbf{q,s}\right) \right] \right\rangle .  \label{sigma-diff}
\end{equation}%
where the dots stand for the remaining contributions of squared amplitudes
from Eq.(\ref{scattering amp}), and the sum is over $j=\mathrm{e}$, MAG,
POL, SO, PV and IND. The formulas to transform this expression to the
laboratory reference frame can be found in Ref. \cite{Turchin/1965}. As long
as the nucleus is free to recoil, the total cross section does not change
and changes in the angular distribution of the scattered neutrons appear
only at order $m_{\mathrm{n}}/m_{\mathrm{A}}$. Since we are not 
interested in angular distributions and will only consider atoms much heavier
than the neutron, we may use the scattering cross section as given above in
Eq.(\ref{sigma-diff}).
\section{Scattering amplitude due to the NLQED induced electric dipole moment}
We now turn to the calculation of the new amplitude $f_{\mathrm{IND}}$ due
to the NLQED induced electric dipole moment $\mathbf{p}_{\mathrm{IND}}$
given in Eq.(\ref{p-ind}). The magnetic moment  of the neutron, $\mathbf{m}$, can be written as
\begin{equation}
\mathbf{m}=\mu _{\mathrm{n}}\mathbf{\sigma },
\end{equation}%
where
\begin{equation}
\mu _{\mathrm{n}}=-9.662\times 10^{-27}\,\mathrm{A\, m}^{2}\;,  \label{mag-mom-n}
\end{equation}
and  $\mathbf{\sigma }=2\mathbf{s}$ are the Pauli matrices. From Eq. (\ref{a-constraint}) there follows the lower bound
\begin{equation}
a>7.6\,\mathrm{fm}.  \label{a}
\end{equation}%
According to Eq.(\ref{pot-energy}), $\mathbf{p}_{\mathrm{IND}}$ interacts
with the atomic electrostatic field, which for simplicity we consider as
given by a pointlike nucleus with electric charge $Ze$,
\begin{equation}
\mathbf{E}_{0}=\frac{1}{4\pi \epsilon _{0}}\frac{Ze}{r^{2}}\mathbf{e}_{r}.
\label{field nucleus}
\end{equation}%
As discussed later, one can neglect electric field shielding due to the atomic
electrons.
Using Eqs.(\ref{field nucleus}), (\ref{Hamiltonian}) and (\ref{f}) one obtains
\begin{equation}
f_{\mathrm{IND}}\left( q,R,a,\beta \right) =\frac{\zeta M\mu _{0}\mu _{%
\mathrm{n}}^{2}Z^{2}e^{2}}{320\pi ^{4}\hbar ^{2}\epsilon _{0}^{3}a^{3}}\left[
-13 \,I_{1}\left( q,R\right) +49 \,I_{2}\left( q,R,\beta \right) \right] ,
\label{f-QED}
\end{equation}%
where $\beta $ is the angle between $\mathbf{s}$ and $\mathbf{q}$. The two
integrals $I_{1}\left( q,R\right) $ and $I_{2}\left( q,R,\beta \right) $ can be easily
 calculated analytically in polar coordinates with $\mathbf{q}$ along the polar axis.
They are given by 
\begin{equation}
I_{1}\left( q,R\right) =\int \frac{\exp \left( i\mathbf{q}\cdot \mathbf{r}%
\right) }{r^{4}}\mathrm{d}^{3}r,  \label{I-1}
\end{equation}%
and 
\begin{equation}
I_{2}\left( q,R,\beta \right) =\int \frac{\exp \left( i\mathbf{q}\cdot 
\mathbf{r}\right) }{r^{4}}\left( \cos \beta \cos \theta +\sin \beta \sin
\theta \cos \varphi \right) ^{2}\mathrm{d}^{3}r.  \label{I-2}
\end{equation}%
The radial integration extends over all space, excluding a sphere of radius $%
R$ around the nucleus. For a heavy nucleus like lead, electric fields as
strong as $10^{23}\,\mathrm{Vm}^{-1}$ exist close to the nuclear surface.
This exceeds by far the critical field, Eq.(\ref{E-crit}), beyond which
higher order terms in the one-loop effective Lagrangian in Eq.(\ref{L-EH})
become important \cite{Dunne} and thus cannot be neglected. Therefore, $R$
has to be much larger than the nuclear radius $R_{\mathrm{N}}$, and we
choose it here as the distance from the nucleus to where the critical field
is reached, i.e.
\begin{equation}
E_{\mathrm{c}} = \frac{1}{4\pi \epsilon _{0}}\frac{Ze}{R^{2}}\;.
\label{critical field condition}
\end{equation}%
For lead isotopes with $Z=82$ one has $R\simeq 300\,\mathrm{fm}$.
The integrals Eqs.(41) and (42) can be solved analytically with the result
\begin{eqnarray}
I_{1}\left( q,R\right)  &=&\frac{2\pi }{R}\left\{ \cos \left( qR\right) +%
\frac{\sin \left( qR\right) }{qR}+\left[ \func{Si}\left( qR\right) -\frac{%
\pi }{2}\right] qR\right\} \bigskip \mathstrut \medskip   \label{I-1-sol} \\%
[0.4cm]
&\simeq &\frac{4\pi }{R}\left[ 1-\frac{\pi }{4}qR+\frac{1}{6}\left(
qR\right) ^{2}-...\right] ,  \nonumber
\end{eqnarray}%
where $\func{Si}\left( x\right) $ is the Sine integral, and
\begin{eqnarray}
I_{2}\left( q,R,\beta \right)  &=&\frac{\pi }{4R}\left\{ \frac{1}{\left(
qR\right) ^{2}}\left[ 2+3\left( qR\right) ^{2}+\left( 6+\left( qR\right)
^{2}\right) \cos \left( 2\beta \right) \right] \cos \left( qR\right) \right. 
\nonumber \\
&&-\frac{1}{\left( qR\right) ^{3}}\left[ 2-3\left( qR\right) ^{2}+\left(
6-\left( qR\right) ^{2}\right) \cos \left( 2\beta \right) \right] \sin
\left( qR\right)   \nonumber \\
&&\left. -qR\left( 3+\cos \left( 2\beta \right) \right) \left( \frac{\pi }{2}%
-\func{Si}\left( qR\right) \right) \right\} \medskip   \label{I-2-sol} \\%
[0.4cm]
&\simeq &\frac{4\pi }{3R}\left[ 1-\frac{3\pi }{32}\left( 3+\cos \left(
2\beta \right) \right) qR+\frac{1}{10}\left( 2+\cos \left( 2\beta \right)
\right) \left( qR\right) ^{2}-...\right] .  \nonumber
\end{eqnarray}%
Using Eqs.(\ref{I-1-sol}) and (\ref{I-2-sol}) in Eq.(\ref{f-QED}) one finds
the final expression for the scattering amplitude due to the NLQED induced
electric dipole moment%
\begin{eqnarray}
f_{\mathrm{IND}}\left( q,R,a,\beta \right)  &=&\frac{\zeta M\mu _{0}\mu _{%
\mathrm{n}}^{2}Z^{2}e^{2}}{320\pi ^{3}\hbar ^{2}\epsilon _{0}^{3}a^{3}R}%
\left\{ \frac{49\left[ 1+3\cos \left( 2\beta \right) \right] \left[ qR\cos
\left( qR\right) -\sin \left( qR\right) \right]}{2\left( qR\right) ^{3}}%
\right.   \label{f-final} \\
&&\left. +\frac{1}{4}\left[ 43+49\cos \left( 2\beta \right) \right] \left[
\cos \left( qR\right) +\frac{\sin \left( qR\right) }{qR}+\left( \func{Si}%
\left( qR\right) -\frac{\pi }{2}\right) qR\right] \right\} \medskip  
\nonumber \\[0.4cm]
&\simeq &\frac{\zeta M\mu _{0}\mu _{\mathrm{n}}^{2}Z^{2}e^{2}}{24\pi
^{3}\hbar ^{2}\epsilon _{0}^{3}a^{3}R}\left\{ 1-\frac{3\pi }{320}\left[
43+49\cos \left( 2\beta \right) \right] qR\right.   \nonumber \\
&&\left. +\frac{1}{100}\left[ 33+49\cos \left( 2\beta \right) \right] \left(
qR\right) ^{2}-...\right\} .  \nonumber
\end{eqnarray}%
The scattering amplitude $f_{\mathrm{IND}}$  exhibits a welcome peculiar dependence on the angle $\beta$ between the neutron spin $\mathbf{s}$ and the three-momentum transfer $\mathbf{q}$. This dependence introduces an asymmetry which for suitable experimental conditions is essentially free from {\it background} contributions due to other well known effects. The largest effect is obtained by evaluating  $f_{\mathrm{IND}}$ at $\beta = 0$ and at $\beta = \pi/2$. This feature will then play a crucial role in the experimental detection of the NLQED induced electric dipole moment of the neutron, as discussed in the following section.
\section{Scattering asymmetry due to the NLQED induced electric dipole moment}
We define the scattering asymmetry as
\begin{equation}
A\left( q,R,a\right) =\frac{\left( \mathrm{d}\sigma /\mathrm{d}\Omega
\right) _{\Vert }-\left( \mathrm{d}\sigma /\mathrm{d}\Omega \right) _{\perp }%
}{\left( \mathrm{d}\sigma /\mathrm{d}\Omega \right) _{\Vert }+\left( \mathrm{%
d}\sigma /\mathrm{d}\Omega \right) _{\perp }},  \label{def-A}
\end{equation}%
where the differential cross section, Eq.(\ref{sigma-diff}), is evaluated for
two neutron polarization states $\mathbf{P}_{\Vert }$ and $\mathbf{P}_{\perp
}$, parallel and perpendicular to the scattering vector $\mathbf{q}$,
respectively. The interference term between the coherent nuclear amplitude
and the amplitude of interest leads  to
\begin{equation}
A_{\mathrm{IND}}\left( q,R,a\right) =\frac{4\pi }{\sigma _{\mathrm{s}}}%
\overline{a_{\mathrm{N}}}\left( f_{\mathrm{IND}\bot }-f_{\mathrm{IND}\Vert
}\right) P,  \label{A-new}
\end{equation}%
where $f_{\mathrm{IND}\Vert }=f_{\mathrm{IND}}\left( q,R,a,0\right) $ and $%
f_{\mathrm{IND}\bot }=f_{\mathrm{IND}}\left( q,R,a,\pi /2\right) $, $%
P=\left\vert \mathbf{P}_{\Vert }\right\vert \,=\left\vert \mathbf{P}_{\perp
}\right\vert $, and $\sigma _{\mathrm{s}}=\sigma _{\mathrm{s,b}} M^{2}/m_{%
\mathrm{n}}^{2}$ is the total scattering cross section of the free atom. We
argue below that in a well-designed experiment possible influences of
interference terms other than between $\overline{a_{\mathrm{N}}}$ and $f_{%
\mathrm{IND}}$ are negligible. Hence, $A\left( q,R,a\right) \simeq A_{%
\mathrm{IND}}\left( q,R,a\right) $, so that the asymmetry defined in Eq.(\ref%
{def-A}) should allow for a detection and determination of the new
amplitude. Using the values for natural lead from Eq.(\ref{values-Pb}) in
Eq.(\ref{A-new}) one obtains
\begin{equation}
A\left( q,R,a\right) \simeq \frac{f_{\mathrm{IND}\bot }-f_{\mathrm{IND}\Vert
}}{9.5\,\mathrm{fm}}P.  \label{A-lead}
\end{equation}%
From Eqs.(\ref{f-final}) and (\ref{A-lead}) it follows that  $%
A\left( q,R,a\right) \propto \chi \left( qR\right) /R$, where $\chi $ is a
function of the dimensionless parameter $qR$. The maximum of $A\left(
q,R,a\right) $ occurs for (see Fig.~1)
\begin{equation}
qR=1.68.  \label{qR}
\end{equation}%
Using the value of $a$ given in Eq.(\ref{a}) the
asymmetry becomes
\begin{equation}
A\left( 5.6\times 10^{12}\,\mathrm{m}^{-1},\,300\,\mathrm{fm},\,7.6\,\mathrm{%
fm}\right) =1.4\times 10^{-3}P,  \label{A-epi}
\end{equation}%
a result which appears experimentally accessible. Neglecting nuclear recoil,
valid in good approximation for neutron scattering off a heavy target, one
may use the relation
\begin{equation}
q=2k\sin \frac{\Theta }{2},
\end{equation}%
where $\Theta $ is the angle  between $\mathbf{k}$ and $\mathbf{k}^{\prime }$, and
$k\equiv\left\vert \mathbf{k}\right\vert =2.197\times 10^{-4}\,\mathrm{fm}%
^{-1}\sqrt{E(\mbox{eV})}$, with $E$ the neutron kinetic energy in $\mathrm{eV}$. In a
backscattering geometry, i.e. for $\Theta \simeq \pi $, the maximum
asymmetry, Eq.(\ref{A-epi}), is obtained with epithermal neutrons of energy $%
E\simeq 165\,\mathrm{eV}$.
\begin{figure}
\centerline{\includegraphics{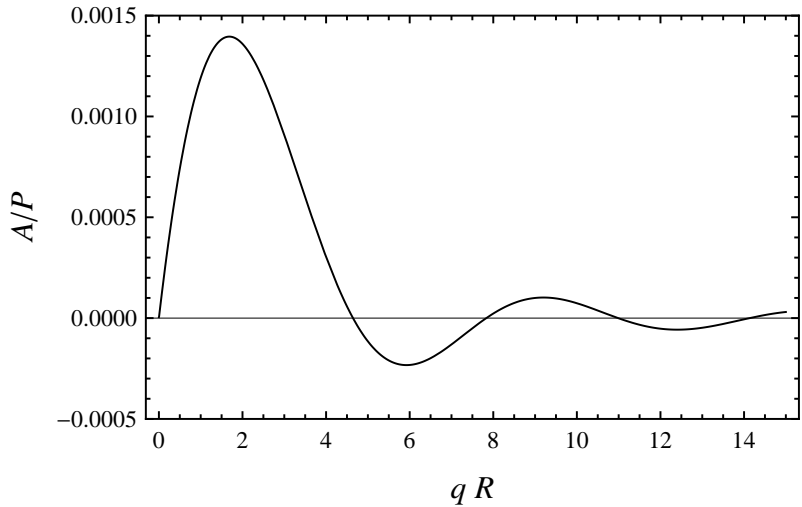}}
\caption{The scattering asymmetry $A\left( q,\,R=300\,\mathrm{fm},\,a=7.6\,\mathrm{fm}%
\right)$, Eq.(\protect\ref%
{A-lead}), normalized to the neutron polarization $P$ as a function of the dimensionless variable 
$qR$.\label{figure1}}
\end{figure}
The result for the asymmetry depends explicitly on the radii $R$ and $a$ of
spheres centered on the nucleus and on the neutron, respectively. These parameters play the role of  separating the
short-distance physics close to the electromagnetic sources from the
long-distance effects involving fields  weak enough for the
Euler-Heisenberg approximation to be valid. In this regard it is important to point out that
electric and magnetic fields induced by quantum fluctuations in the QED
vacuum involve various multipolarities \cite{Dominguez/2009/1}. For
instance, in the case of a neutron in an external, quasistatic electric
field $\left\vert \mathbf{E}_{0}\right\vert <E_{\mathrm{c}}$, the induced
electric field Eq.(\ref{E-induced}) has a dipole type term of order 
$\mathcal{O}(\left\vert \mathbf{x}\right\vert ^{-3})$, as well as a higher
multipole of order $\mathcal{O}(\left\vert \mathbf{x}\right\vert ^{-6})$,
while the induced magnetic field involves terms of order $\mathcal{O}%
(\left\vert \mathbf{x}\right\vert ^{-3})$, $\mathcal{O}(\left\vert \mathbf{x}%
\right\vert ^{-5})$, and $\mathcal{O}(\left\vert \mathbf{x}\right\vert ^{-9})
$, with $\mathbf{x=r}-\mathbf{r}_{\mathrm{n}}$. These higher order
multipoles can be safely neglected in the interaction energy as long as the
weak-field approximation remains valid. However, it is not clear what happens at
distances closer to the nucleus or to the neutron. This problem is similar
to that of the separation into far and near field regions around a localized
charge/current distribution in classical electrodynamics, where the
lowest-order multipole provides the long-distance solution. Although one
cannot compute the interaction energy due to QED vacuum effects stemming
from the regions $r<R$ and $\left\vert \mathbf{r}-\mathbf{r}_{\mathrm{n}%
}\right\vert <a$, their contribution to the asymmetry would presumably have a
different dependence on $\mathbf{s}$ and $\mathbf{q}$. For large $q$ it
might smear out the oscillations appearing in Fig.~\ref{figure1}. For small $q\,$, corresponding to small spatial resolution in probing the
QED vacuum, the asymmetry should not be affected much by the
short-distance physics. This statement is underlined by the fact that at
leading order in $qR$ the $\beta $-dependent term in $f_{%
\mathrm{IND}}$, Eq.(\ref{f-final}), does not depend on $R$. For $q R\ll 1$%
, one thus obtain a prediction for the asymmetry which should be robust
against variations in the choice of $R$, i.e.
\begin{equation}
A\left( q\ll R^{-1},R,a\right) \simeq A\left( q,a\right) =\frac{49}{320}%
\frac{\overline{a_{\mathrm{N}}}}{\sigma _{\mathrm{s}}}\frac{\zeta M\mu
_{0}\mu _{\mathrm{n}}^{2}Z^{2}e^{2}}{\pi \hbar ^{2}\epsilon _{0}^{3}a^{3}}q P \;.
\label{A-small-q}
\end{equation}%
For a neutron energy of $1\,\mathrm{eV}$ in a backscattering geometry one
expects
\begin{equation}
A\left( 4.4\times 10^{11}\,\mathrm{m}^{-1},\,7.6\,\mathrm{fm}\right)
=2.2\times 10^{-4}P \;.  \label{A-hot}
\end{equation}
From a practical point of view the polarization of
epithermal neutrons with more than $100\,\mathrm{eV}$ requires a spin filter
of polarized protons. This is technically demanding if one wishes to
polarize a beam with a diameter of several cm. The measured energy-dependent
neutron polarization cross section of a polarized-proton spin filter for the
energy range of interest may be found in Ref. \cite{Lushchikov/1970}. On the
other hand, neutrons with energies up to $\sim 1\,\mathrm{eV}$ are available
from a hot moderator in a reactor neutron source with much higher intensity
than epithermal neutrons. They can be polarized using magnetic monochromator
crystals, or by a spin filter of polarized $^{3}$He gas \cite{Heil/1998}
which has a polarization cross section proportional to $k^{-1}$. For the
fluxes available at the Institut Laue-Langevin (ILL) in Grenoble an
asymmetry as in Eq.(\ref{A-hot}) appears experimentally accessible within a
few days of beam time.\\
To conclude this section we stress that the neutron scattering asymmetry due
to nonlinear QED has two characteristic properties which should make it
rather easy to detect  and distinguish  from other effects. First, the
asymmetry attains its maximum value for perpendicular orientations of the
neutron polarization and vanishes for opposite orientations. This is in
contrast to most ordinary asymmetries which become maximal for opposite
orientations. Second, $A_{\mathrm{IND}}\left( q,R,a\right) $ exhibits a
characteristic $q$ dependence with a broad maximum around the value of $q$
given in Eq.(\ref{qR}). These features are discussed in more detail in the sequel.
\section{Analysis of background asymmetries}
In this section we study the contributions to the asymmetry $%
A\left( q,R,a\right) $, Eq.(\ref{def-A}), from the various ordinary scattering
amplitudes defined in Eq.(\ref{scattering amp}). The neutron
spin-dependent amplitudes can be written as
\begin{equation}
f\left( \mathbf{q,s}\right) =f_{0}\left( \mathbf{q}\right) +f_{1}\left( 
\mathbf{q}\right) \left[ \mathbf{s\cdot w}\left( \mathbf{q}\right) \right] 
\label{f-spin-dep}
\end{equation}%
where $f_0\left( \mathbf{q}\right)$ is spin independent, and $\mathbf{w}$ is a vector not correlated with the neutron spin. In the case of the
weak amplitude $f_{\mathrm{PV}}$ the vector $\mathbf{q}$ must be replaced
by $\mathbf{k}$.
For instance, for the nuclear amplitude in Eq.(\ref{a-spin}%
), $\mathbf{w}$ is independent of $\mathbf{q}$ and given by the nuclear spin 
$\mathbf{I}$. It can be shown in general that the terms proportional to $%
\langle \left( \mathbf{s}\cdot \mathbf{w}\right) ^{2}\rangle $ in the
differential cross section, Eq.(\ref{sigma-diff}), are all independent of
the neutron polarization and therefore cannot generate an asymmetry. In
principle these terms influence the size of $A\left( q,R,a\right) $ through
the total scattering cross section $\sigma _{\mathrm{s}}$, Eq.(\ref{A-new}). For scattering angles $\Theta \rightarrow 0$ the pure spin-orbit
cross section, quadratic in the amplitude $f_{\mathrm{SO}}$, might become large enough
to have  an impact on $\sigma _{\mathrm{s}}$. However, for sufficiently large $%
\Theta $, and in the absence of nuclear and electronic polarization,
corrections to $\sigma _{\mathrm{s}}$ due to squared-amplitude terms can be
safely  neglected.
The interference terms  between the nuclear and the
other scattering amplitudes in Eq.(\ref{sigma-diff}) may however affect the asymmetry $A\left(
q,R,a\right) $ through their dependence on the neutron polarization. This requires careful consideration, and we start with the amplitude $f_{\mathrm{SO}%
}$ for spin-orbit scattering. It originates in the interaction of the
neutron magnetic moment with the magnetic field present in the neutron rest
frame due to its motion through the atomic electric fields. Its expression
is (see e.g. \cite{Squires/1978})
\begin{equation}
f_{\mathrm{SO}}\left( \mathbf{q,s}\right) =i\frac{M}{m_{\mathrm{n}}}\cot
\left( \Theta /2\right) \frac{\mu _{\mathrm{n}}\mu _{0}}{2\pi \hbar }%
eZ\left[ 1-F\left( q\right) \right] \left( \mathbf{s}\cdot \mathbf{n}\right)
,  \label{f-so}
\end{equation}%
where $eZ\left[ 1-F\left( q\right) \right] $ is the Fourier transform of the
electric charge density of the atom. This term involves the nuclear charge $Z
$ and the atomic form factor $F\left( q\right) $ normalized to $F\left(
0\right) =1$. This form factor is measured e.g. in X-ray scattering off atoms, and is a real function of the momentum.
The unit vector $\mathbf{n}$ points along $\mathbf{k\times k}^{\prime }$, so
that the amplitude can contribute only if the neutron polarization has a
component out of the scattering plane. The asymmetry due to the spin-orbit
interaction is given by
\begin{equation}
A_{\mathrm{SO}}=\frac{\func{Im}\;\overline{a_{\mathrm{N}}}}{\sigma _{\mathrm{s}%
}}\frac{M}{m_{\mathrm{n}}}\cot \left( \Theta /2\right) \frac{\mu _{\mathrm{n}%
}\mu _{0}}{\hbar }eZ\left[ 1-F\left( q\right) \right] \left( \mathbf{P}%
_{\Vert }\cdot \mathbf{n}-\mathbf{P}_{\perp }\cdot \mathbf{n}\right) .
\label{A-so}
\end{equation}%
The imaginary part of $\overline{a_{\mathrm{N}}}$ above is due to nuclear
absorption and can be calculated using the optical theorem. Lead nuclei
absorb neutrons only weakly so that $\func{Im}\; \overline{a_{\mathrm{N}}}%
\simeq \left( 4\pi \right) ^{-1}k\sigma _{\mathrm{s}}$ for neutron energies
of interest here. For a scattering angle $\Theta =\pi /2$ a neutron kinetic
energy of $E\simeq 330\,\mathrm{eV}$ would be required to observe the
maximum asymmetry according to Eq.(\ref{A-epi}). In this case the atomic
form factor $F\left( q\right) \simeq 0$, and a maximum asymmetry $A_{\mathrm{%
SO}}$ should be observed for $\mathbf{P}_{\perp }$ perpendicular to the
scattering plane, i.e. $\mathbf{P}_{\perp }\cdot \mathbf{n}=P$ (while $%
\mathbf{P}_{\Vert }\cdot \mathbf{n}=0$ by definition). As a result, for $%
E=330\,\mathrm{eV}$ one has  $A_{\mathrm{SO}}\simeq 4.8\times 10^{-4}P$, while for $%
E=1\,\mathrm{eV}$ together with the conservative value $F\left( q\right) =0$%
, the asymmetry becomes $A_{\mathrm{SO}}\simeq 2.6\times 10^{-5}P$. These
values are already smaller than the corresponding values of $A_{\mathrm{IND}%
}$, but with suitable experimental settings they can  be reduced even further. Notice that the condition $\mathbf{P}_{\perp }\cdot \mathbf{q}=0$
required for $f_{\mathrm{IND}\perp }$ can be realized for different
orientations of $\mathbf{P}_{\perp }$, while also fulfilling $%
\mathbf{P}_{\perp }\cdot \mathbf{n}=0$. Hence, choosing a
backscattering geometry, for which $\cot \left( \Theta /2\right) \ll 1$ together with
$\mathbf{P}_{\perp }\cdot \mathbf{n}\simeq 0$, and using realistic
assumptions about the experimental definition of the directions of $\mathbf{q
}$ and $\mathbf{P}$, one can easily suppress $A_{\mathrm{SO}}$ by a factor $
50$. Hence, the impact of $A_{\mathrm{SO}}$ can be kept well under the $1 \%$ level.\\
Next, the amplitude $a_{\mathrm{N}}$ of the neutron-nuclear interaction in
Eq.(\ref{a-spin}), when squared, gives rise to interference between its spin-dependent and
spin-independent parts. After ensemble averaging this becomes proportional to $P$ and
to the nuclear polarization $P_{\mathrm{N}}$. In thermal equilibrium, $P_{%
\mathrm{N}}=\tanh \left( \mu _{\mathrm{N}}B/\left( k_{\mathrm{B}}T\right)
\right) $ for nuclei with magnetic moment $\mu _{\mathrm{N}}$ in a magnetic
field $B$ ($k_{\mathrm{B}}$ is the Boltzmann constant). If the target is at
room temperature, and given that no magnetic field is needed at the position
of the sample, the $P$-dependent cross section is orders of magnitude too
small to have an impact on the asymmetry.\\
We consider next  the amplitude $f_{\mathrm{MAG}}$, which is due to the
interaction of the neutron magnetic moment with the magnetic field produced
by unpaired atomic electrons of paramagnetic contaminants in the sample. The
operator structure of this amplitude is given by 
\begin{equation}
f_{\mathrm{MAG}}\propto \mathbf{s}\cdot \left[ \mathbf{e}_{q}\times \mathbf{M%
}\left( \mathbf{q}\right) \times \mathbf{e}_{q}\right] ,  \label{f-mag}
\end{equation}%
where $\mathbf{M}\left( \mathbf{q}\right) $ is the Fourier transform of the
total (spin and orbital) magnetization of the atom, and $\mathbf{e}_{q}=%
\mathbf{q}/q$. The interference term with $\overline{a_{\mathrm{N}}}$ thus
involves the neutron polarization and the sample-averaged magnetization. It
would only influence the asymmetry if (1) the ratio $B/T$ is sufficiently
high to result in a sizable magnetization, (2) paramagnetic centers are
sufficiently abundant, (3) the two neutron polarization states in the
asymmetry have different projections perpendicular to the scattering plane
(which is a consequence of the term in brackets in Eq.(\ref{f-mag})), and
(4) measurements are performed for sufficiently small $q$, where the
magnetic form factor still has a sizable value. Regarding the latter, even
for the smaller value of $q$ envisaged in Eq.(\ref{A-hot}), the magnetic
form factor leads to a strong suppression of $f_{\mathrm{MAG}}$. Hence, with
the conditions 1, 2 and 3 under experimental control one can safely disregard
magnetism as a source of an asymmetry.\\
Finally, the parity-violating amplitude $f_{\mathrm{PV}}$ due to the
hadronic weak interaction may lead to a different type of asymmetry which
has indeed been observed in neutron transmission experiments. Effects depend
on the neutron helicity, hence $\mathbf{w}=\mathbf{k}$ in Eq.(\ref%
{f-spin-dep}), with a complex coefficient to describe both parity violating
spin rotation and transmission asymmetry. The amplitude is normally so small
that it requires special efforts to detect it. For thermal neutrons,
transmission asymmetries for longitudinally polarized neutrons \cite%
{Forte/1980} have typical sizes of a few times $10^{-6}$. However, for neutron
energies in the vicinity of p-wave resonances of complex nuclei a strong
enhancement due to the weak nuclear interactions may appear. A prominent
example is the transmission asymmetry of $7\%$ found at the p-wave resonance
of $0.76\,\mathrm{eV}$ in $^{139}$La \cite{Alfimenkov/1983}. However, no
effect sufficiently strong to affect the asymmetry $A_{\mathrm{IND}}$ is
known for lead in the relevant energy range. In addition, an experimental
test can easily be performed. In fact, taking the neutron polarization $%
\mathbf{P}$ parallel and anti-parallel to $\mathbf{q}$, i.e. $\beta =0$ and $%
\beta =\pi $, it follows from Eqs.(\ref{f-final}) and (\ref{A-new}) that $%
A_{\mathrm{IND}}=0$ for these two polarization orientations. In contrast,
for $f_{\mathrm{PV}}$ one has $A_{\mathrm{PV}}\propto \sin \left( \Theta
/2\right) $, which could be measured separately and corrected for if the need
arises.
\section{Comparison of the NLQED  amplitude with ordinary electric
amplitudes}
The electric amplitudes $f_{\mathrm{POL}}$ and $f_{\mathrm{e}}$  do not
generate any known scattering asymmetry. However, owing to their
characteristic $q$-dependences a comparison with the NLQED amplitude $f_{%
\mathrm{IND}}$ is needed. Like $f_{\mathrm{IND}}$, the amplitude $%
f_{\mathrm{POL}}$ due to the electric polarizability 
of the neutron, $\alpha _{\mathrm{n}}$,  is induced by the nuclear electric field, Eq.(\ref{field
nucleus}). In SI units $\alpha _{\mathrm{n}}$ is defined by $\mathbf{p}=4\pi
\epsilon _{0}\alpha _{\mathrm{n}}\mathbf{E}_{0}$, so that its dimension is $\left[ \alpha _{%
\mathrm{n}}\right] =\mathrm{m}^{3}$. The calculation of $f_{\mathrm{POL}}$
follows from  Eq.(\ref{f}) with the interaction energy given by Eq.(\ref{H-el-pol}).
This leads to the integral  Eq.(\ref{I-1}) with the result given in Eq.(\ref%
{I-1-sol}). However, the lower limit of the radial integration is now
different from that for $f_{\mathrm{IND}}$. In fact, this lower limit can now be extended
down to the nuclear radius $R_{\mathrm{N}}$, since for $r>R_{\mathrm{N}}$
the neutron probes only the long-range electric forces. For $r<R_{\mathrm{N}}
$ the electric interaction is small in comparison with the nuclear force, so
that in early calculations \cite{Barashenkov/1957}-\cite{Thaler/1959} it has simply
been included in the nuclear amplitude. In SI units and for the electric
field given in Eq.(\ref{field nucleus}), the dependence of $f_{\mathrm{POL}}$
on $q$ is given by 
\begin{equation}
f_{\mathrm{POL}}\left( q\right) \simeq \frac{1}{4\pi \epsilon _{0}}\frac{M}{%
\hbar ^{2}}\frac{Z^{2}e^{2}}{R_{\mathrm{N}}}\alpha _{\mathrm{n}}\left\{ 1-%
\frac{\pi }{4}qR_{\mathrm{N}}+\frac{1}{6}\left( qR_{\mathrm{N}}\right)
^{2}-...\right\} .  \label{f-QCD}
\end{equation}%
The term linear in $q$ is characteristic of the $r^{-4}$ dependence of the
Hamiltonian, and it also enters the interference term in the cross section.
This feature has been exploited in the past to measure $\alpha _{\mathrm{n}}$. Conflicting results from experiments performed during more than three
decades show that a proper assessment of all systematic errors has been
difficult (see e.g. the table in Ref. \cite{Schmiedmayer/1989}). The most
recent result \cite{Schmiedmayer/1991}, derived from energy-dependent
neutron transmission through a $^{208}$Pb target, and reporting the smallest
uncertainty is
\begin{equation}
\alpha _{\mathrm{n,}\exp }=\left( 1.20\pm 0.15\pm 0.20\right) \times
10^{-3}\,\mathrm{fm}^{3}.  \label{alpha-exp}
\end{equation}%
Calculations using quark bag models \cite{Bernard/1988} agree with this
result. An early estimate of Breit and Rustgi \cite{Breit/1959} using data
on pion photoproduction   already indicated that $\alpha _{\mathrm{n}%
}<2\times 10^{-3}\,\mathrm{fm}^{3}$. These authors also analyzed other
effects which might mimic a signal from the neutron electric polarizability.
From an estimate of  vacuum polarization effects close to the
nucleus, and using the Uehling potential \cite{Uehling/1935}, they concluded
that this contribution to neutron scattering can be safely  neglected. 
Turning now to the question of shielding of the nuclear charge by the atomic
electrons, one notices that this might quench the amplitude $f_{\mathrm{IND}}
$. Since $f_{\mathrm{POL}}$ and $f_{\mathrm{IND}}$ both depend quadratically
on the electric field $\mathbf{E}_{0}$, one can draw parallels with the
analysis of $f_{\mathrm{POL}}$. We recall that in our calculation of $f_{%
\mathrm{IND}}$ one needs to exclude a spherical region of radius $R$ around
the nucleus inside which the weak-field expansion of the Euler-Heisenberg
Lagrangian breaks down. This procedure was followed (for different reasons)
in the early calculations of $f_{\mathrm{POL}}$ \cite
{Barashenkov/1957}-\cite{Thaler/1959}, \cite{Breit/1959} where the nuclear region with
radius $R_{\mathrm{N}}$ was excluded. A more recent analysis which does not
rely on a simple model for the nuclear charge distribution 
gives \cite{Sears/1986}
\begin{equation}
f_{\mathrm{POL}}\left( q\rightarrow 0\right) =\frac{1}{4\pi \epsilon _{0}}%
\sqrt{\frac{3}{\pi }}\frac{M}{\hbar ^{2}}\frac{Z^{2}e^{2}}{r_{\mathrm{N}}}%
\alpha _{\mathrm{n}},  \label{f-pol-q-to-0}
\end{equation}%
where $r_{\mathrm{N}}$ is the root mean square charge radius of the nucleus.
This shows that slow neutrons are indeed insensitive to details at this
length scale. Notice that this result is nearly identical to the leading term
in Eq.(\ref{f-QCD}) after replacing $R_{\mathrm{N}}$ by $r_{\mathrm{N}}$. In
the derivation of Eq.(\ref{f-pol-q-to-0}) the following intermediate result
was obtained in \cite{Sears/1986} 
\begin{equation}
f_{\mathrm{POL}}\left( q\rightarrow 0\right) \propto \int_{0}^{\infty
}\left\vert F_{\mathrm{N}}\left( \kappa \right) -F\left( \kappa \right)
\right\vert ^{2}\mathrm{d}\kappa =\int_{0}^{\infty }\left\vert F_{\mathrm{N}%
}\left( \kappa \right) \right\vert ^{2}\mathrm{d}\kappa \left[ 1-\mathcal{O}%
\left( R_{\mathrm{N}}/R_{\mathrm{A}}\right) \right] ,  \label{f with F}
\end{equation}
where $F_{\mathrm{N}}\left( \kappa \right) $ and $F\left( \kappa \right) $
are the charge form factors of the nucleus and of the electron distribution
in the atom, respectively. With $R_{\mathrm{N}}/R_{\mathrm{A}}\simeq 10^{-5}$, shielding
of the nuclear charge can be neglected in $f_{\mathrm{POL}}$. Even in the
limit $q\rightarrow 0$ the neutron feels the full unscreened nuclear charge
as far as the electric polarizability is concerned. For the NLQED induced
electric dipole moment it follows that, with $R\simeq 300\,\mathrm{fm}$ for
lead, the corresponding correction term of order $\mathcal{O}\left( R/R_{\mathrm{A}%
}\right) $ is much larger than the one of order $\mathcal{O}\left( R_{\mathrm{N}}/R_{\mathrm{A%
}}\right) $. However, $R/R_{\mathrm{A}}\lesssim 10^{-2}$ is still small
enough so that one can neglect shielding of the electric field in the region around the
nucleus.
We now compare the two amplitudes $f_{\mathrm{POL}}$ and $f_{\mathrm{IND}}$
in the limit $q\rightarrow 0$, i.e. their respective contributions to the
neutron scattering length. Setting $R_{\mathrm{N}}=1.2\,\mathrm{fm} \, A^{1/3}$
and using Eq.(\ref{alpha-exp}) one can  estimate the leading order term in
Eq.(\ref{f-QCD}) as
\begin{equation}
f_{\mathrm{POL}}\left( q\rightarrow 0\right) \simeq 0.04\,\mathrm{fm}\,.
\label{f-pol-q0}
\end{equation}%
From Eq.(\ref{f-final}) the corresponding leading term in $f_{\mathrm{IND}}$ is
\begin{equation}
f_{\mathrm{IND}}\left(q \rightarrow 0\right) =\frac{1}{4\pi \epsilon _{0}}
\frac{M}{\hbar ^{2}}\frac{Z^{2}e^{2}}{R}\alpha _{\mathrm{IND}}\;,
\end{equation}
where
\begin{equation}
\alpha _{\mathrm{IND}}=\frac{\zeta \mu _{0}\mu _{\mathrm{n}}^{2}}{6\pi
^{2}\epsilon _{0}^{2}a^{3}}\simeq \frac{3.3\,\mathrm{fm}^{3}}{\left( a\left[ 
\mathrm{fm}\right] \right) ^{3}}\,.
\end{equation}
With the value of $a$ from Eq.(\ref{a-constraint}) one obtains
\begin{equation}
\alpha _{\mathrm{IND}}=7.5\times 10^{-3}\,\mathrm{fm},
\end{equation}
which is larger than $\alpha _{\mathrm{n}}$, Eq.(\ref{alpha-exp}). However, 
the impact of this polarizability on the amplitude $f_{\mathrm{IND}}$
for $q \rightarrow 0$ is  suppressed with
respect to $f_{\mathrm{POL}}$  due to $R\gg R_{\mathrm{N}}$. In fact, for lead one
obtains
\begin{equation}
f_{\mathrm{IND}}\left(q \rightarrow 0\right) \simeq 0.006\,\mathrm{fm}\;.
\end{equation}
This result, though, is  not even  an order of magnitude smaller than $f_{%
\mathrm{POL}}\left( q\rightarrow 0\right) $, Eq.(\ref{f-pol-q0}). It would
thus contribute about $5\times 10^{-11}\,\mathrm{eV}$ to the neutron optical
potential of solid lead, which is quite substantial given the high precision of some
neutron optical methods. The contribution of the NLQED electric dipole moment to the total cross
section is $\sigma _{\mathrm{IND}}\simeq -8\pi \overline{a}f_{\mathrm{IND}%
}\left( q\rightarrow 0\right) $, where   $\overline{a_{%
\mathrm{N}}}\simeq \overline{a}$ in Eq.(\ref{sigma-diff}) has been used. Numerically this becomes 
\begin{equation}
\sigma _{\mathrm{IND}}\left( q\rightarrow 0\right) =-0.014\times 10^{-24}%
\mathrm{cm}^{2}.
\end{equation}
\begin{figure}
\centerline{\includegraphics{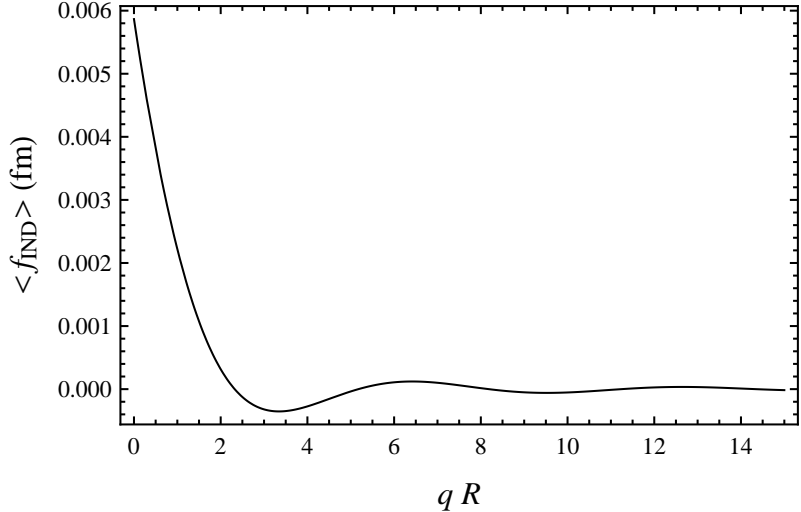}}
\caption{Amplitude $\left\langle f_{\mathrm{IND}}\left( q, R=300\,
\mathrm{fm}, a=7.6\,\mathrm{fm}\right) \right\rangle $ for unpolarized
neutrons, Eq.(\protect\ref{f_IND-av}), as a function of the dimensionless variable 
$qR$.\label{figure2}}
\end{figure}
We  discuss next the amplitude $f_{\mathrm{e}}$ which describes the
interaction of the electric charges of the atom with the internal charge
distribution of the neutron as characterized by its mean squared charge
radius. The amplitude for a bound nucleus is given by $f_{\mathrm{e}}=-b_{%
\mathrm{e}}Z\left[ 1-F\left( q\right) \right] $, with the atomic form factor 
$F\left( q\right) $ given in Eqs.(\ref{f-so}) and (\ref{f with F}), and the
neutron-electron scattering length is $b_{\mathrm{e}}\simeq -1.35\times
10^{-3}\,\mathrm{fm}$ as determined from measurements of the total cross
sections of lead and bismuth at different neutron energies \cite
{Koester/1976}-\cite{Kopecky/1997}. For lead and sufficiently large $q$, $f_{%
\mathrm{e}}=0.11\,\mathrm{fm}$, which leads to a contribution to the total scattering cross
section
 \cite%
{Sears/1986} $\sigma _{\mathrm{e}}\simeq -8\pi \overline{a}b_{\mathrm{e}%
}Z\simeq 0.25\times 10^{-24}\,\mathrm{cm}^{2}$. It is also interesting to notice that since $f_{\mathrm{e}}\left(
q\rightarrow 0\right) \rightarrow 0$ this amplitude does not contribute to
the neutron scattering length. The atomic form factor changes significantly
at small $q$  where interference of neutron waves from different
atoms cannot be neglected. Hence, the macroscopic state of the sample
 enters crucially in the analysis of scattering data. In contrast, in the case
of $f_{\mathrm{IND}}$ where larger effects show up at much higher values
of $q$, interatomic interferences do not play any significant role.\\
To conclude this section we discuss the contribution of $f_{\mathrm{IND}}$
as a potential background in measurements of the amplitudes $f_{\mathrm{POL}}
$ and $f_{\mathrm{e}}$ performed with unpolarized neutrons. For the
polarization averaged $f_{\mathrm{IND}}$ one obtains
\begin{eqnarray}
\left\langle f_{\mathrm{IND}}\left( q,R,a\right) \right\rangle  &=&\frac{1}{2%
}\int_{0}^{\pi }f_{\mathrm{IND}}\left( q,R,a,\beta \right)\, \sin \beta \;
\mathrm{d}\beta   \nonumber \\
&=&\frac{\zeta M\mu _{0}\mu _{\mathrm{n}}^{2}Z^{2}e^{2}}{48\pi ^{3}\hbar
^{2}\epsilon _{0}^{3}a^{3}R}\left\{ \cos \left( qR\right) +\frac{\sin \left(
qR\right) }{qR}+\left[ \func{Si}\left( qR\right) -\frac{\pi }{2}\right]
qR\right\}   \label{f_IND-av}
\end{eqnarray}%
To analyze the
impact on $f_{\mathrm{e}}$ one may approximate relativistic Hartree-Fock
results for $F\left( q\right) $ by the simple function $\left[ 1+3\left(
q/q_{0}\right) ^{2}\right] ^{-1/2}$ \cite{Sears/1986}. With $F\left(
q_{0}\right) =1/2$ the momentum $q_{0}$ provides a scale at which significant changes in
$f_{\mathrm{e}}$ take place. For lead, $q_{0}=8.3\times 10^{10}\,\mathrm{m}%
^{-1}$, and
\begin{equation}
f_{\mathrm{e}}\left( q_{0}\right) -f_{\mathrm{e}}\left( 0\right) =50\times
10^{-3}\,\mathrm{fm}.
\end{equation}%
In contrast, the change of $\left\langle f_{\mathrm{IND}}\right\rangle $ is
much smaller due to its milder $q$-dependence and its smaller magnitude (see
Fig.~\ref{figure2}),
\begin{equation}
\left\langle f_{\mathrm{IND}}\left( q_{0}\right) \right\rangle -\left\langle
f_{\mathrm{IND}}\left( 0\right) \right\rangle =-0.11\times 10^{-3}\,\mathrm{%
fm}.
\end{equation}%
Since the precision of the best  measurements of $b_{\mathrm{e}}$ is at the
level of a few percent, potential background due to $\left\langle f_{\mathrm{IND}%
}\right\rangle $ is negligible. A similar argument leads to the same
conclusion for the determination of $\alpha _{\mathrm{n}}$ from $f_{\mathrm{%
POL}}$. 
\section{Conclusions}
Many, if not most proposals to detect nonlinear effects due to quantum
fluctuations in the QED vacuum rely on experiments involving lasers of ultra-high
intensities \cite{Schwinger/1951}-\cite{Fried/1966}. These intensities, though, are at least
two orders of magnitude below current values. An alternative approach has
been discussed in this paper, based on the theoretical prediction of an
induced electric dipole moment of the neutron, ${\mathbf{p}}_{\mathrm{IND}}$%
, in an external quasistatic electric field \cite{Dominguez/2009}. The
peculiar features of this dipole moment, particularly its dependence on the
angle between ${\mathbf{p}}_{\mathrm{IND}}$ and the neutron spin, suggests
the definition of an asymmetry which could be detected in the scattering of
polarized neutrons from heavy nuclei. We have introduced this asymmetry and
discussed all possible sources of background asymmetries. We have also
compared the new NLQED amplitude with  ordinary electric scattering
amplitudes, particularly the one due to the polarization of the neutron in an
electric field due to its quark substructure. The conclusion from
this detailed analysis is that the asymmetry due to NLQED should be
observable using epithermal neutrons, and even using thermalized neutrons from a
hot moderator. This would be the first ever experimental confirmation of
nonlinearity in electrodynamics due to QED vacuum fluctuations.
The numerical predictions for the asymmetry made in this paper were calculated using
definite values for the parameters $R$ and $a$. These  were derived
from the condition that the electric and magnetic fields  should be below their critical values, beyond which the weak-field expansion of the effective Lagrangian breaks down. While the
value of the asymmetry $A$ for small $q$ does not depend on $R$, it 
does depend on  $a$ as seen from Eq.(\ref{Hamiltonian}) Hence,
the numerical results given here should be correct up to a numerical factor of order one.
\section{Acknowledgments}
This work was supported in part by FONDECYT 1095217 (Chile), Proyecto
Anillos ACT119 (Chile), by CONICET (PIP 01787) (Argentina), ANPCyT (PICT
00909)(Argentina), and UNLP (Proy.~11/X492) (Argentina), NRF (South Africa),
and National Institute for Theoretical Physics (South Africa).

\end{document}